\documentclass[aps,prb,a4paper,twocolumn,showpacs,showkeys,fleqn,floatfix,superscriptaddress]{revtex4-2}

\usepackage[dvips]{epsfig}
\usepackage{color}
\usepackage{amsmath}
\usepackage{multirow}

\newcommand{\figwidth}{\columnwidth}

\newcommand{\vect}[1]{\mathbf{#1}}
\newcommand{\bamn}{Ba$_2$MnGe$_2$O$_7$}

\begin{document}

\title{Magnetic resonance in the quasi-2D square lattice easy-plane antiferromagnet \bamn{}}

\author{V.~N.~Glazkov}
\email{glazkov@kapitza.ras.ru}
\affiliation{P. L. Kapitza Institute for Physical Problems RAS, Kosygin str. 2, 119334 Moscow, Russia}

\author{Yu.~V.~Krasnikova}
\affiliation{P. L. Kapitza Institute for Physical Problems RAS, Kosygin str. 2, 119334 Moscow, Russia}

\author{I.~K.~Rodygina}
\affiliation{P. L. Kapitza Institute for Physical Problems RAS, Kosygin str. 2, 119334 Moscow, Russia}

\author{H.-A.~Krug von Nidda}
\affiliation{Experimental Physics V, University of Augsburg, Augsburg, Germany}

\author{T.~Masuda}
\affiliation{Institute for Solid State Physics, The University of Tokyo, Kashiwa, Chiba 277-8581, Japan}

\begin{abstract}
We report results of a multi-frequency (0.8-60~GHz) electron spin resonance study of the spin dynamics in the quasi-2D square lattice antiferromagnet \bamn{} both in antiferromagnetically ordered and paramagnetic phases. We directly observe two zero-field gaps in the excitation spectrum of the ordered phase, the larger one being due to easy-plane anisotropy, and the smaller one indicates the presence of  fourth-order in-plane anisotropy probably related to the multiferroic properties of this compound. We observe effects of hyperfine interaction on the electron spin resonance spectra in the antiferromagnetically ordered state, which turns out to be comparable with in-plane anisotropy. The hyperfine field strength is found from the observed low-temperature electron spin resonance data. The spin dynamics of the paramagnetic phase is characterized by strong broadening of the ESR absorption line, which can be ascribed to the vortex dynamics of a 2D magnet.
\end{abstract}

\maketitle

\section{Introduction}

Low-dimensional magnetic systems are actively studied during the last decades \cite{dejong,vasiliev}. This interest is grounded on the key role of quantum and thermal fluctuations in these systems which suppress conventional magnetic ordering and allow for the formation of a spin-liquid state in a low-dimensional magnet. In the case of two dimensional (2D) systems the magnetic properties are strongly dependent on the presence of anisotropic interactions. While a Heisenberg 2D magnet orders at $T=0$ only, Ising anisotropy leads to the formation of the ordered state at finite temperature (see, e.g., \cite{matis}), while planar XY-anisotropy results in the topological Berezinskii-Kosterlitz-Thouless (BKT) transition \cite{bkt,QMC-XY-BKT}. As all real magnets are three-dimensional, the interplay of the anisotropic interactions and the weaker inter-chain or inter-plane couplings can lead to the possible appearance of new phases \cite{fortune,zhitomirskii}. Multiferroicity (the combination of magnetic order and electric polarization allowing to control electric properties via magnetic field and vice versa) is another emergent topic of magnetism \cite{mostovoy,ratcliff,pyatakov} which further enriches the phase diagram of many low-dimensional magnets.
Finally, magnetic dynamics of the coupled electronic and nuclear spin subsystems is a problem of interest with a long history. This coupling gives rise to  nuclear spin waves and affects the electronic spin waves by increasing (or even creating) a gap in the electronic spin-wave spectrum \cite{heeger,degennes,witt,borovik,andrienko,andrienkoproz1,andrienkoproz2,zaliznyakzorin,prsos}.

In the present paper we report the results of the magnetic resonance study of the Mn-based ($S=5/2$) 2D square-lattice antiferromagnet \bamn{}. This compound stays on the crossroad of the above mentioned concepts allowing to test these effects and their possible interplay. It orders antiferromagnetically at $T_{\rm N}=4.0$~K as confirmed by elastic and inelastic neutron scattering and thermodynamic measurements \cite{masuda2010}. Neutron diffraction suggests easy-plane ordering with the spins confined to the $(001)$ plane of the tetragonal crystal and aligned along the $[100]$ direction at zero field \cite{masuda2010}.  In-plane and inter-plane exchange integrals were determined from the dispersion curves as $J_{||}=27.8$ $\mu$eV and $J_\perp=1.0$ $\mu$eV, correspondingly \cite{masuda2010}. Magnetization curves \cite{masuda2010} demonstrate saturation at the field of 97.5~kOe at $\vect{H}||[110]$ and a weak feature at 1.2~kOe at the same field orientation interpreted as traces of a spin-flop transition. An accurate recent magnetization and neutron scattering study \cite{zaliznyak} refined the transition field value as approximately 0.6~kOe and reports its unusual temperature dependence. Instability of magnons at high fields (approx 0.8 of the saturation value) was observed in a part of the Brillouin zone in agreement with the predictions of Ref.~\cite{zhitomirskii2}. \bamn{} is a multiferroic material \cite{murakawa,sazonov,iguchi}, where weak magnetic field dependent electric polarization up to 1 $\mu$C/m$^2$ appears for $\vect{H}||[110]$ \cite{murakawa}. The magnetic resonance study of Ref.~\cite{iguchi} revealed the presence of  magnon gap of approximately 26~GHz (0.11~meV) and an unusual effect of nonreciprocity of microwave transmission through the sample due to the multiferroic properties of \bamn{}.
Recent neutron scattering experiments with extreme resolution in near-backscattering geometry \cite{zaliznyak} revealed the presence of a smaller gap of 0.036~meV (about 9~GHz) at 50~mK with unusual temperature dependence governed by hyperfine interaction and electric polarization of the multiferroic state.

We have performed a detailed electron spin resonance (ESR) study of \bamn{} down to 0.45~K. This study confirms the collinear antiferromagnetic ordering pattern below the N\'{e}el temperature. A continuous shift of the ESR absorption spectra down to the lowest temperatures indicates the important role of the hyperfine coupling in the spin dynamics of this compound. We have also observed anisotropy of the resonance absorption in the plane normal to the main axis, which is explained in terms of fourth order anisotropy. Broadening of the ESR line above the N\'{e}el temperature bears traces of a vortex relaxation mechanism related to the BKT transition of an ideal planar 2D magnet.

\section{Experimental details and samples}

We use samples from the same batch as the samples used in Ref.~\cite{masuda2010}. \bamn{} crystallizes into tetragonal space group $D_{\rm2d}^3$ with the lattice parameters $a=8.505${\AA} and $c=5.528${\AA} \cite{sazonov}. Each Mn$^{2+}$ ion is located in the center of oxigen tetrahedron which is slightly constrained along the $[001]$ direction.

The directions of the crystallographic axes were checked by X-ray diffraction on a Bruker APEX II diffractometer and small single crystals (from 4 to 160 mg) were cut from the oriented sample to be used in different experimental setups.

ESR spectra at frequencies from 0.8 GHz to 100 GHz were recorded at P.Kapitza Institute (Moscow, Russia) using a set of the home-made transmission-type spectrometers. The spectrometers operating at the frequencies $f>9$~GHz use rectangular or cylinder microwave cavities. The low frequency ESR spectrometer ($f<5$~GHz) uses specially designed toroidal microwave resonators.  All spectrometers were equipped with cryomagnets with the maximal field up to 120~kOe. One of the spectrometers was equipped with a  He-3 vapor pumping cryostat with the base temperature of 0.45~K.

Accurate X-band (9.40 GHz) ESR measurements of temperature (4 to 300~K) and angular dependencies of ESR absorption were performed at the University of Augsburg  (Augsburg, Germany) with a Bruker Elexys  spectrometer equipped with a helium  flow cryostat and programmable goniometer.

\section{Experimental results}
\subsection{Temperature dependencies of resonance absorption}
\begin{figure}
\epsfig{clip=, width=\figwidth,file=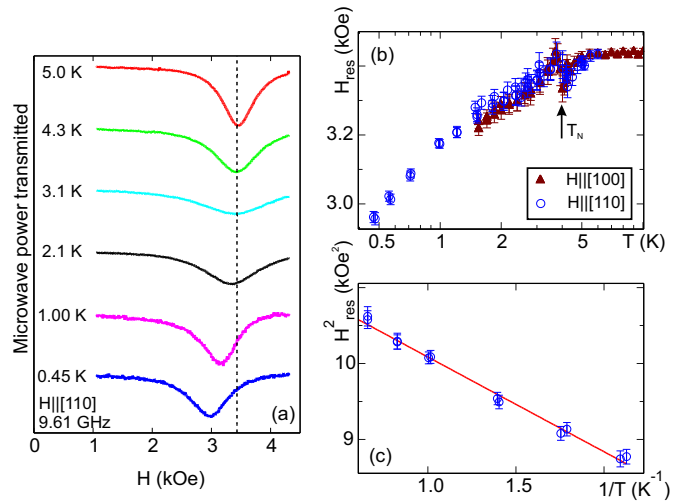}
\caption{(color online) (a) Temperature evolution of the  ESR absorption at $\vect{H}||[110]$ down to 0.45~K, $f=9.61$~GHz. (b) Temperature dependence of the resonance field for $\vect{H}||[110]$ (circles) and $\vect{H}||[100]$ (filled triangles), $f=9.61$~GHz. Arrow marks the N\'{e}el temperature. (c) Resonance field squared as a function of inverse temperature for $\vect{H}||[110]$, symbols -- experiment, solid line -- model of hyperfine pulling (Eq.~(\ref{eqn:linfit})), $f=9.61$~GHz. }
\label{fig:he3(t)}
\end{figure}

\begin{figure}
\epsfig{clip=, width=\figwidth,file=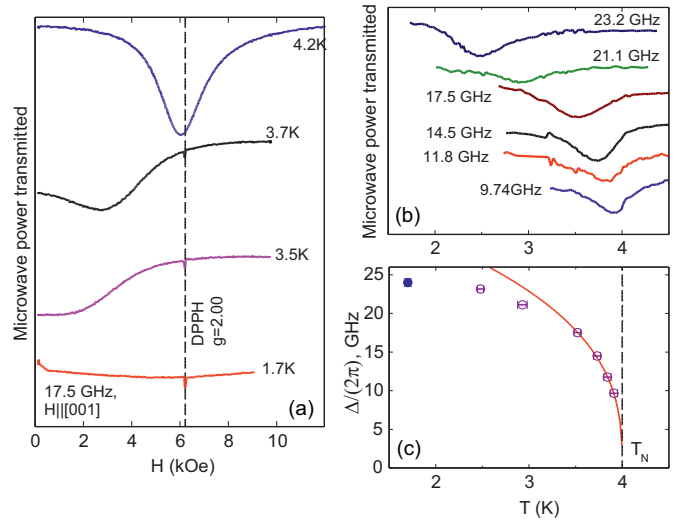}
\caption{(color online) (a) Temperature evolution of the ESR absorption for the field $\vect{H}||[001]$, $f=17.5$~GHz, $T\leq 4.2$~K. Vertical dashed line marks the position of the paramagnetic resonance with $g=2.00$, the narrow absorption line at this field is a DPPH marker. (b) Zero-field ESR absorption as a function of temperature (``temperature resonance'') at different microwave frequencies. (c) Temperature dependence of the magnon gap. Symbols -- experimental data determined from the ``temperature resonance'' experiment (open circles) and determined from $f(H)$ dependencies at $T=1.7$~K (filled circle); curve -- phenomenological fit $\Delta\propto (1-T/T_{\rm N})^\beta$ with exponent $\beta\approx0.36$, vertical dashed line marks the N\'{e}el temperature. }
\label{fig:scans!!c}
\end{figure}

At high temperatures ($T>8$~K) we observe an ESR absorption signal with $g=2.00\pm0.02$ for all field orientations, which is typical for the $S=5/2$ $L=0$ Mn$^{2+}$ ions.  As temperature approaches the N\'{e}el point ($T_{\rm N}=4.0$~K) a weak shift of the resonance line to  lower fields is systematically observed below approximately 7~K both for $\vect{H}||[001]$ and for $\vect{H}\perp[001]$ (Fig.~\ref{fig:he3(t)}).   The temperature range $T_{\rm N}<T<7$~K, where the resonance field shift is observed, is the same as the temperature range where the broad maximum of the magnetic susceptibility typical for low-dimensional magnets is observed in Ref.\cite{masuda2010}. This resonance field shift above the N\'{e}el point is isotropic within 1\% accuracy of our experiment and can be ascribed to the formation of short range correlations in a quasi-2D magnet.

On cooling below the transition temperature we observe an anisotropic shift of the ESR absorption. At $\vect{H}||[001]$ the resonance field strongly decreases on cooling (see Fig.~\ref{fig:scans!!c}). This behavior corresponds to the opening of the energy gap in the magnon spectrum. For an easy-plane antiferromagnet the antiferromagnetic resonance frequency (which corresponds to the energy of $q=0$ magnons) for the field applied along the hard axis is given by the equation \cite{goorevich}
\begin{equation}\label{eqn:gapped}
\omega=\sqrt{(\gamma H)^2+\Delta^2},
\end{equation}

\noindent here $\gamma$ is the gyromagnetic ratio and $\Delta$ is the temperature dependent gap.

The temperature dependence of the energy gap $\Delta$ can be followed by a so called ``temperature resonance'' experiment: we measure the temperature dependence of the microwave power absorbed by the sample as a function of temperature at zero field (Fig.~\ref{fig:scans!!c}). The maximal absorption (minimum of the transmitted microwave power) is observed when the gap in the magnon spectrum is equal to the microwave frequency used. The resulting temperature dependence is shown at Fig.~\ref{fig:scans!!c}, where data close to the N\'{e}el point can be phenomenologically fitted as $\Delta=\Delta_0 \left(1-T/T_{\rm N} \right)^\beta$ with $\beta=0.36\pm0.04$. Within the mean-field model the gap in the magnon spectrum is proportional to the order parameter \cite{kubo}, the obtained critical exponent is close to the  critical exponent of the order parameter for the 3D Ising model $\beta_{\rm Ising}^{\rm (3D)}\approx0.327$ \cite{kolesik} and for the 3D XY-model $\beta_{\rm XY}^{\rm (3D)}=0.3485$ \cite{campostrini}.

At $\vect{H}\perp[001]$ the change of the absorption spectrum below the N\'{e}el point is by far less dramatic. Resonance absorption is observed down to the lowest temperature of 0.45~K (Fig.~\ref{fig:he3(t)}). We have found that at microwave frequencies $f>9$~GHz the resonance absorption shifts to lower fields both for $\vect{H}||[100]$ and $\vect{H}||[110]$. Moreover, this temperature dependent shift continues monotonously down to 0.45~K ($\simeq T_{\rm N}/8$) and does not saturate,  while the temperature dependence of the gap $\Delta$ saturates at approximately $T_{\rm N}/2$.

\subsection{Angular dependencies of resonance absorption in the $(001)$ plane}
\begin{figure}
\epsfig{clip=, width=\figwidth,file=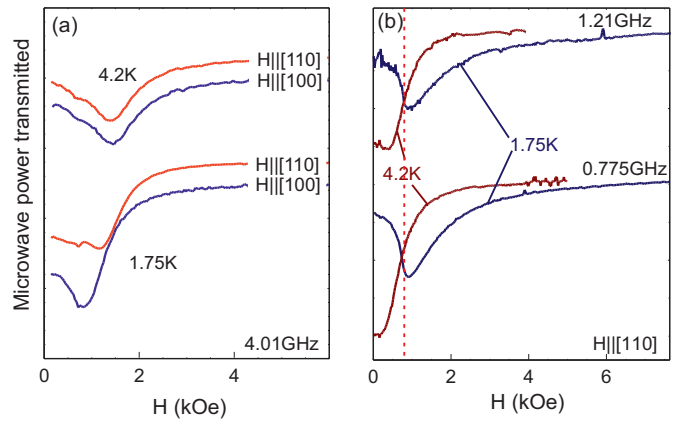}
\caption{(color online) Examples of ESR absorption spectra at low frequencies $f<5$~GHz: (a) Comparison of $\vect{H}||[110]$ and $\vect{H}||[100]$ ESR absorption at $f=4.01$~GHz. (b) Temperature evolution of $\vect{H}||[110]$ ESR absorption at $f<1.5$~GHz. The vertical dashed line indicates the estimate of the spin-reorientation field $H_{\rm sr}=0.78$~kOe from the asymmetric edge of the low-temperature absorption.  }
\label{fig:low_freqs}
\end{figure}
\begin{figure}
\epsfig{clip=, width=\figwidth,file=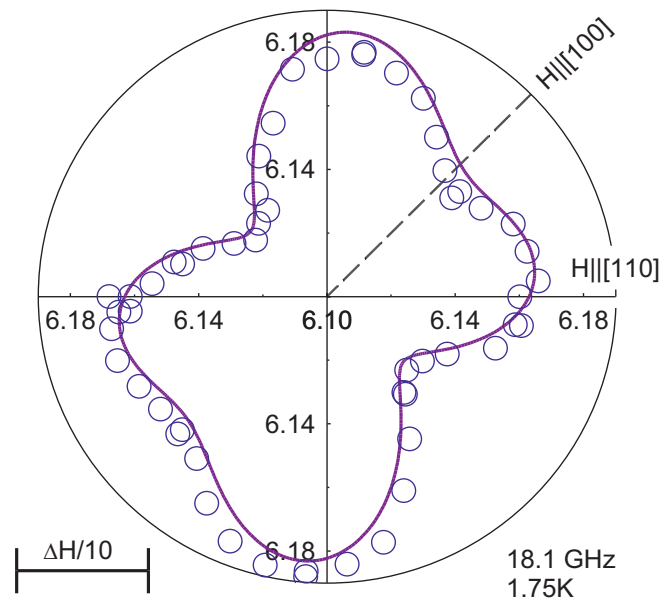}
\caption{(color online) $(001)$-plane angular dependence (polar plot) of the AFMR field  at $T=1.75$~K, $f=18.1$~GHz. Circles -- experimental data,  solid line -- fit by the sum of 2-nd and 4-th harmonics (see Eq.~(\ref{eqn:angularfit})). Horizontal bar shows 1/10 of the ESR linewidth which is a conventional estimate of the accuracy of resonance field determination. }
\label{fig:angular}
\end{figure}

The magnetic response of the ideal easy-plane antiferromagnet should be isotropic within the easy plane. However, magnetization measurements of Refs.~\cite{masuda2010,zaliznyak} demonstrate anisotropy of magnetization within the $(001)$ plane. Near-backscattering geometry inelastic neutron scattering experiments \cite{zaliznyak} revealed the presence of the minute gap in the excitation spectrum equal to 0.036~meV (about 9~GHz) at 50~mK related to this in-plane anisotropy.

Antiferromagnetic resonance (AFMR) curves also demonstrate the presence of the in-plane anisotropy: at microwave frequencies $f>9$~GHz  the  $\vect{H}||[110]$ resonance field is systematically slightly higher (on the edge of experimental error) than the $\vect{H}||[100]$ resonance field.

At lower frequencies ($f<5$~GHz) the temperature evolution of ESR absorption at $\vect{H}\perp[001])$ became clearly anisotropic (Fig.~\ref{fig:low_freqs}). At $\vect{H}||[100]$ the absorption line strongly shifts to lower fields below $T_{\rm N}$, while for $\vect{H}||[110]$ this shift is either much smaller (at 4.01 GHz) or the  AFMR absorption shifts to higher fields (at 1.21 and 0.778 GHz). The shape of the resonance absorption curve at the lowermost frequencies is asymmetric: it features a rather sharp left shoulder at $(0.78\pm0.07)$~kOe. Such asymmetry is sometimes observed close to a spin-reorientation transition, when the ESR eigenfrequency sharply changes at a certain field. However, interpretation of the low-frequency data requires particular attention because  the linewidth in the frequency domain is comparable with the microwave frequency: e.g., at 4.01~GHz (Fig.~\ref{fig:low_freqs}) half-linewidth of AFMR absorption at 1.7K is about 600~Oe, which corresponds to approx. 1.5~GHz half-linewidth in the frequency domain. Thus, it becomes difficult to discern ``true'' resonance absorption when the field scan at a fixed frequency crosses the $f(H)$ dependence for one of the eigenmodes from non-resonant absorption when the experimental scan passes through the wing of the resonance line only.

Additionally, we have accurately measured  the angular dependence of the resonance field for $\vect{H}\perp[001]$ at 18.1~GHz (Fig.~\ref{fig:angular}). We observed a regular modulation of the resonance field value, which can be fitted by the sum of second and fourth harmonics:

\begin{equation}
\frac{H}{H_0}=
1+
A_2\cos\left(2\left(\varphi-\xi_2\right)\right)+
A_4\cos\left(4\left(\varphi-\xi_4\right)\right)
,
\label{eqn:angularfit}
\end{equation}

\noindent here $A_2$ and $A_4$ are the relative amplitudes of the harmonics, $\xi_{1,2}$ describe arbitrary misorientation of the sample. Best fit values are $H_0=(6.159\pm0.001)$~kOe, $A_2=-(0.0027\pm0.0002)$, $A_4=(0.0022\pm0.0001)$, $\xi_2=-(27\pm2)^0$, $\xi_4=(0.7\pm1.1)^0$. The second harmonic is probably related to a small (few degrees) deviation of the sample rotation axis from the $[001]$ axis and the fourth harmonic describes the anisotropy of the resonance field within the $(001)$ plane.

\subsection{Frequency-field diagrams}
\begin{figure}
\epsfig{clip=, width=\figwidth,file=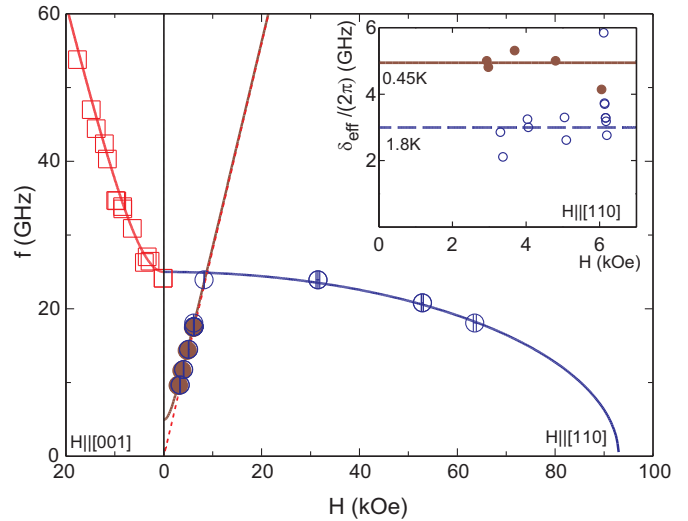}
\caption{(color online) Frequency-field diagram for antiferromagnetic resonance in \bamn{} at $f>9$~GHz. Squares -- $\vect{H}||[001]$, $T=1.75$~K; open circles -- $\vect{H}||[110]$, $T=1.8$~K; closed circles -- $\vect{H}||[110]$, $T=0.45$~K. Solid lines -- model curves, as described in the text, dashed line -- $f=\gamma H$. Inset: effective gap $\delta_{\rm eff}$ for $\vect{H}||[110]$ at different temperatures.}
\label{fig:f(h)}
\end{figure}

At $f>9$~GHz the measured frequency-field diagrams for \bamn{} (Fig.~\ref{fig:f(h)}) practically follow the known dependencies for the collinear easy-plane antiferromagnet \cite{goorevich,kubo,iguchi}. For the field applied along the main symmetry axis $[001]$ we observe a gapped mode which can be fitted by Eq.~(\ref{eqn:gapped}) with zero-field gap $\Delta/(2\pi)=(25\pm 1)$~GHz at 1.8~K and gyromagnetic ratio $\gamma/(2\pi)=2.80$~GHz/kOe.

For the field applied perpendicular to the symmetry axis the theory of antiferromagnetic resonance in an easy-plane antiferromagnet \cite{goorevich} predicts at low fields one field-independent mode with the same gap $\Delta$, which softens  at the saturation field, and one gapless mode with $\omega=\gamma H$. However, there is a systematic temperature dependent deviation from the expected linear behavior  at $\vect{H}\perp[001]$. The observed decrease of the resonance field can be phenomenologically interpreted as  opening of the second smaller gap

\begin{equation}
\delta_{\rm eff}=\sqrt{\omega^2-\left(\gamma H\right)^2}.
\label{eqn:deltaeff}
\end{equation}

\noindent This transformation (see inset to Fig.~\ref{fig:f(h)}) demonstrates that the effective gap is temperature dependent and $\delta_{\rm eff}$ varies from 3~GHz at 1.8~K to 5~GHz at 0.45~K. These values agree well with inelastic neutron scattering results of Ref.~\cite{zaliznyak}.

\subsection{High temperature ESR linewidth temperature and angular dependencies}

\begin{figure}
  \centering
  \epsfig{clip=, width=\figwidth,file=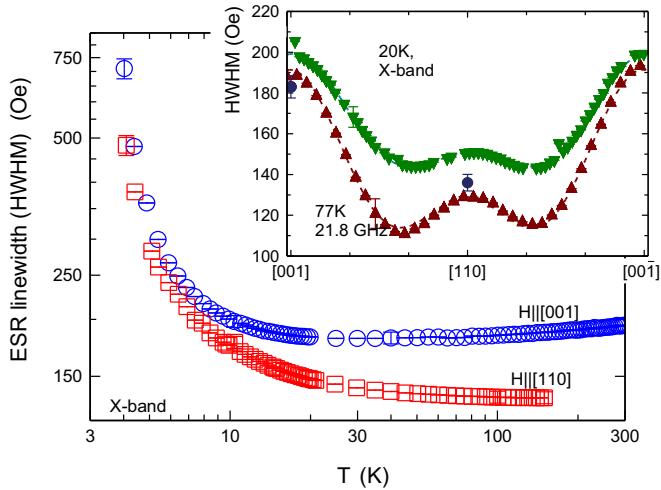}
  \caption{(color online) Main panel: Temperature dependence of the EPR linewidth (half-width at half-maximum) for $\vect{H}||[001]$ (circles) and $\vect{H}||[110]$  (squares) measured with Bruker X-band (9.40~GHz) spectrometer. Inset: angular dependencies of EPR linewidth at 20~K (measured with Bruker X-band spectrometer) and at 77~K (measured at 21.8~GHz with transmission type spectrometer), filled circles show linewidth values for $\vect{H}||[100],[110]$ at 77~K from temperature dependencies  of linewidth measured with Bruker X-band (9.40~GHz) spectrometer (see main panel).}\label{fig:width_exp}
\end{figure}

The anisotropic behavior below the N\'{e}el temperature reflects the presence of anisotropic spin-spin interactions in \bamn{}. The same interactions are known to affect spin relaxation in the paramagnetic phase thus determining the ESR linewidth. To access this ``alter ego'' of anisotropic interactions we have measured temperature and angular dependencies of the ESR parameters. These data are shown in Fig.~\ref{fig:width_exp}.

The linewidth for $\vect{H}||[001]$ exceeds that for $\vect{H}||[110]$ over the broad temperature range from room temperature down to the N\'{e}el point. The ESR line broadens on cooling below  30~K for both field directions indicating some sort of critical behavior. Above 30~K the temperature dependence of the ESR linewidth is qualitatively different for different field directions: for $\vect{H}||[110]$ the ESR linewidth decreases on heating approaching some high-temperature limit, while for $\vect{H}||[001]$ the ESR linewidth slightly increases on heating from 182~Oe at 30~K to 192~Oe at 300K.

The angular dependence of the ESR linewidth demonstrates similar modulations at 20~K and 77~K. For the $\vect{H}||[001]$ to $\vect{H}||[110]$ field rotation the maximal linewidth is observed at $\vect{H}||[001]$, minimal linewidth is observed for the field applied at approx. $60^\circ$ from the $[001]$ axis. The amplitude of the linewidth modulation is temperature dependent.

\section{Discussion}
\subsection{Fourth-order in-plane anisotropy effects}
\begin{figure}
\epsfig{clip=, width=\figwidth,file=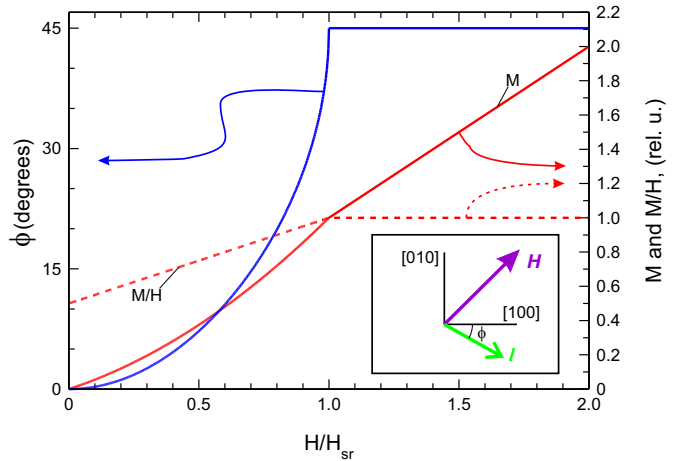}
\caption{(color online) Angle of rotation of the order parameter (left axis) and magnetization process (right axis) for the  easy-plane antiferromagnet with in-plane anisotropy. Magnetic field applied along $[110]$ direction, at zero field the order parameter is aligned along $[100]$ direction. Scheme at the inset shows angle measurement convention.}
\label{fig:sr}
\end{figure}
Tetragonal symmetry of \bamn{} results in almost axial symmetry of its magnetic properties. Nevertheless our experiments (Figs.~\ref{fig:low_freqs},~\ref{fig:angular}) clearly demonstrate inequivalence of the $[100]$ and $[110]$ in-plane directions.  Moreover, magnetization measurements of Refs.~\cite{masuda2010,zaliznyak} demonstrate the anisotropy of the low-field susceptibility within  the $(001)$ plane and a spin-reorientation transition at approximately 0.6~kOe (value at $T=2$~K) \cite{zaliznyak} at $\vect{H}||[110]$.

As a first step of our analysis we will consider the effects of fourth-order in-plane anisotropy on magnetization and antiferromagnetic resonance. We will follow  the hydrodynamic approach of Ref.~\cite{andrmar}. This approach describes the eigenmodes of a collinear antiferromagnet at $T=0$ and at small magnetic field as oscillations of the antiferromagnetic order parameter vector described by the Lagrangian density:

\begin{equation}\label{eqn:am-lagr}
    {\cal L}=\frac{\chi_0}{2\gamma^2} \left(\dot{\vect{l}}+\gamma \left[\vect{l}\times\vect{H} \right] \right)^2-U_{\rm A}
\end{equation}

\noindent here  $\vect{l}$ is the antiferromagnetic order parameter (unit vector aligned along the sublattice magnetization in the case of collinear antiferromagnetic order), $\chi_0$ is the transverse static susceptibility, $\gamma$ is the gyromagnetic ratio and $U_{\rm A}$  is the anisotropy energy. The anisotropy energy in the case of fourfold axis can be written as

\begin{equation}\label{eqn:anis4}
    U_{\rm A}=\frac{b}{2} l_z^2+\alpha l_x^2 l_y^2
\end{equation}

\noindent here $z||[001]$, $x||[100]$ and $y||[010]$, $b>0$ is the main anisotropy constant providing strong easy-plane anisotropy and the term with $\alpha$ describes in-plane anisotropy, $\alpha>0$ corresponds to the preferred direction $\vect{l}||[100]$ or $\vect{l}||[010]$. The preferred orientation of the order parameter of \bamn{} along a $[100]$-like direction was demonstrated in neutron scattering experiments \cite{masuda2010} and, as we will demonstrate below, agrees with antiferromagnetic resonance data.

Assuming $\vect{l}||[100]$ at zero field, for $\vect{H}||[010]$ and $[001]$ the equilibrium orientation of the order parameter will remain unchanged. The case of $\vect{l}||[100]$ and $\vect{H}||[100]$ corresponds to the  unfavorable Zeeman energy and becomes unstable at high fields $H>\sqrt{2\alpha/\chi_0}$ with the order parameter jumping to $\vect{l}||[010]$ orientation. Since $[100]$ and $[010]$ orientation are equivalent for the fourfold symmetry, the order parameter would remain in ``flopped'' $\vect{l}||[010]$ orientation even when the field is turned off.

In the case of $\vect{H}||[110]$ the magnetic field is aligned at 45$^\circ$ with respect to the initial orientation of the order parameter and order parameter slowly rotates within the $(001)$ plane to gain magnetization energy. Competition of the magnetization energy and anisotropy energy results in the steady rotation of the order parameter by the angle

\begin{equation}\label{eqn:angle}
    \varphi=\frac{1}{2} \arcsin \frac{\chi_0 H^2}{2\alpha}
\end{equation}

\noindent this rotation stops at the spin-reorientation field

\begin{equation}
\label{eqn:Hsr}
H_{\rm sr}=\sqrt{2\alpha/\chi_0},
\end{equation}

\noindent when the order parameter becomes orthogonal to the magnetic field. As the field is turned off, the order parameter would return to its original orientation. Steady  rotation of the sublattices makes magnetization process at $\vect{H}||[110]$ nonlinear (see also Fig.~\ref{fig:sr}). The longitudinal with respect to the field  magnetization component  reads:

\begin{equation}\label{fig:M(H)}
M/H=(\chi_0/2)(1+H/H_{\rm sr}).
 \end{equation}

Taking for the spin reorientation field a value of $0.7$~kOe (which provides the best description of antiferromagnetic resonance data, as will be explained below) we can compare the experimentally measured (at $T=2$~K and  $H=100$~Oe) \cite{masuda2010} ratio $\chi_{[110]}/\chi_{[001]}\approx0.69$ with the prediction of Eq.~(\ref{fig:M(H)}) $(1/2)((1+H/H_{\rm sr})\approx 0.57$. The values are reasonably close keeping in mind that the magnetization is measured at $\approx T_{\rm N}/2$ when the longitudinal susceptibility of the antiferromagnet is still significant.

The dynamic equations can be found from Eq.~(\ref{eqn:am-lagr}) via the Euler-Lagrange equations. The eigenfrequencies of the order parameter oscillations will feature two zero-field gaps: a larger gap corresponding to the strong easy-plane anisotropy  $\Delta=\gamma\sqrt{b/\chi_0}$ and a smaller gap corresponding to the in-plane anisotropy $\delta=\gamma\sqrt{2\alpha/\chi_0}=\gamma H_{\rm sr}$. The field dependencies of the eigenfrequencies are given as follows:

\noindent $\vect{H}||[001]$ :
  \begin{eqnarray}
  \omega_1&=&\sqrt{\Delta^2+(\gamma H)^2}\nonumber\\
  \omega_2&=&\delta\label{eqn:4-1}
  \end{eqnarray}
  $\vect{H}||[100]$:
   \begin{eqnarray}
  \omega_1&=&\Delta\nonumber\\
  \omega_2&=&\sqrt{\delta^2+(\gamma H)^2}\label{eqn:4-2}
  \end{eqnarray}
$\vect{H}||[110]$, $H<H_{\rm sr}$:
  \begin{equation}
        \left|
    \begin{array}{cc}
    \left[\omega^2-\Delta^2+\right.&2 \imath\omega \gamma H {\cal B}\\
    \left.+\delta^2\left(\frac{1}{2}-{\cal B}^2\right)\right]&\\
    &\\
    -2 \imath\omega \gamma H {\cal B}&\left[\omega^2-\gamma^2 H^2\sin 2\varphi -\right.\\
    &\left.\delta^2 \cos 4\varphi\right]\\
    \end{array}
    \right|=0\label{eqn:4-3}
  \end{equation}
  here  the angle of the order parameter rotation $\varphi$ is defined by Eq.~(\ref{eqn:angle}) and ${\cal B}=(\cos\varphi-\sin\varphi)/\sqrt{2}$ for short.

\noindent $\vect{H}||[110]$, $H>H_{\rm sr}$:
  \begin{eqnarray}
  \omega_1&=&\sqrt{\Delta^2-\delta^2/2}\nonumber\\
  \omega_2&=&\sqrt{-\delta^2+(\gamma H)^2}\label{eqn:4-4}
  \end{eqnarray}

The softening of the $\omega(H)$  dependence close to $H_{\rm sr}$ is asymmetric above and below the spin-reorientation field. This is in agreement with the observed asymmetry of the low-frequency absorption curves (Fig.\ref{fig:low_freqs}) and allows to estimate the spin-reorientation field as the position of the sharp left shoulder of these absorption curves at approximately $(0.78\pm 0.07)$~kOe.

To complete this discussion, we notice that the theoretical model \cite{andrmar} is applicable at small fields $H\ll H_{\rm sat}$ ($H_{\rm sat}$ is the saturation field). From mean-field theory approach \cite{goorevich} we expect that the  field-independent modes of Eqs.~(\ref{eqn:4-1}),(\ref{eqn:4-2}),(\ref{eqn:4-4}) would soften at $H_{\rm sat}$ being proportional at high fields to $\sqrt{1-(H/H_{\rm sat})^2}$.  The best fit of our data yields $H_{\rm sat}=(94\pm2)$~kOe in agreement with the known magnetization study \cite{masuda2010}.

Note here, that there is a qualitative disagreement of this theory with our results. Eq.~(\ref{eqn:4-4}) predicts, that at $\vect{H}||[110]$ resonance field $H_{\rm res}^{[110]}>\omega/\gamma$ (i.e., the resonance absorption shifts to the right from paramagnetic position), while our experiment (Figs.~\ref{fig:he3(t)},~\ref{fig:f(h)}) demonstrates a shift of the ESR absorption to the left from the paramagnetic position. We will demonstrate below that this effect can be ascribed to the influence of hyperfine interaction.

\subsection{Electron-nuclear coupling  and antiferromagnetic resonance in \bamn{}}
\begin{figure}
\epsfig{clip=, width=\figwidth,file=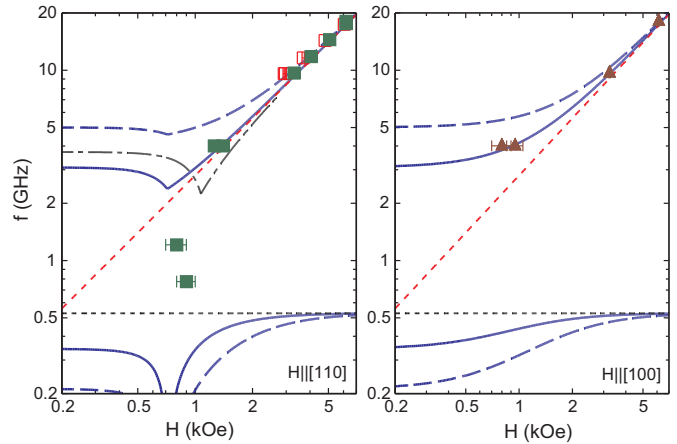}
\caption{(color online) Low-frequency ($f<20$~GHz) frequency-field diagrams for antiferromagnetic resonance in  \bamn{}. Symbols: experimental data, closed symbols: $T=1.75$~K, open symbols on left panel: $T=0.45$~K. Dotted lines show paramagnetic resonance position and unbiased NMR frequency $\omega_{\rm n}/(2\pi)$. Solid curves: model curve for $\delta=2$~GHz and $\omega_{\rm n}/(2\pi)=530$~MHz at $T=1.75$~K. Dashed curves: similar modeling for $\delta=2$~GHz and $\omega_{\rm n}/(2\pi)=530$~MHz at $T=0.45$~K. Dash-dotted curve on left panel shows modeling results  for $\delta=3$~GHz and $\omega_{\rm n}/(2\pi)=500$~MHz at $T=1.75$~K for comparison.}
\label{fig:f(h)low}
\end{figure}

Eqs.~(\ref{eqn:4-2}) and (\ref{eqn:4-4}) predict that antiferromagnetic resonance absorption at $\vect{H}||[100]$ and $\vect{H}||[110]$ should be observed, correspondingly, to the left and to the right from the resonance field in the paramagnetic phase. However, the experiment (Fig.~\ref{fig:he3(t)}) indicates that the resonance field in both orientations shifts to the left from the paramagnetic position and the effect of anisotropy is just a minor difference of these resonance fields. We also recall here that the temperature evolution of the resonance field at $\vect{H}\perp[001]$ continues down to the very low temperatures well below the temperature of order parameter saturation, i.e. $T<T_\textrm{N}/2$.

Such behavior is typical for Mn-based antiferromagnets because of the large hyperfine coupling constant for  $^{55}$Mn nuclei \cite{heeger,degennes,witt,borovik,andrienko,andrienkoproz1,andrienkoproz2,zaliznyakzorin,prsos}. Strong coupling of the electron spin resonance (ESR) and nuclear spin resonance (NMR) modes results in the formation of cooperative electron-nuclear excitations and in the pulling of ESR and NMR modes.

In the regime where the frequency of the ESR mode is much larger than that of NMR (which holds for the  typical ESR frequencies $f>9$~GHz) the effect of hyperfine coupling could be qualitatively understood as  additional effective anisotropy field  caused by nuclear spins polarized by the ordered electronic magnetic moments.  Since the nuclear subsystem is far from saturation, this effective anisotropy field is proportional to the paramagnetic susceptibility of the nuclear subsystem and hence is proportional to $1/T$ and the induced gap is proportional to $1/\sqrt{T}$ .

The temperature dependent smaller gap  at $\vect{H}\perp[001]$ arises due  to this effect. To check it we plot the squared resonance field as a function of inverse temperature in the range of low $T$ (Fig.~\ref{fig:he3(t)}), which fits well to the expected  dependence

\begin{equation}\label{eqn:linfit}
H_{\rm res}^2=H_0^2-a/T
\end{equation}

\noindent with parameters $a=(1.25\pm0.10)  ~ $K$\cdot$kOe${}^2$ and $H_0=(3.37\pm0.03)$~kOe for $f=9.61$~GHz.

The quantitative description of hyperfine pulling of ESR modes in \bamn{} is complicated by the presence of fourth-order in-plane anisotropy which turns out to be comparable in strength with the hyperfine coupling. Hence, both effects should be accounted for simultaneously. We will analyze the effects of hyperfine coupling within the same hydrodynamical approach \cite{andrmar} following the model of Refs.~\cite{udalov,prsos}.  The electronic spin system is considered to be fully saturated (which corresponds approximately to $T<T_{\rm N}/2$). The coupling of the nuclear and electronic spins adds a new term to the potential energy density in the Lagrangian (\ref{eqn:am-lagr}):

\begin{equation}\label{eqn:hf-u-dens}
    U_{\rm HF}=-A \left(\vect{m}_1\vect{l}-\vect{m}_2\vect{l}\right)
\end{equation}

\noindent here $\vect{m}_{1,2}$ denote the magnetization per nucleus for ${}^{55}$Mn nuclei belonging to different sublattices, $\vect{l}$ is the antiferromagnetic order parameter and $A$ is the scaled hyperfine constant.

The effective field acting on each manganese nucleus is $\vect{H}^{({\rm (eff)})}_{1,2}=\pm  (2A/\rho)\vect{l}$, here $\rho$ is the density of manganese atoms in \bamn{}. The equilibrium nuclear polarization equals $\vect{m}=\chi_n \vect{H}^{({\rm (eff)})}$, here $\chi_n=(\gamma_n^2\hbar^2 I (I+1))/(3 k_B T)$ is the paramagnetic nuclear susceptibility per nucleus with the ${}^{55}$Mn nuclear gyromagnetic ratio $\gamma_n/(2\pi)=1.06$~MHz/kOe.

The entire problem now involves three coupled vector equations describing the precession of two nuclear magnetic moments and the precession of the antiferromagnetic order parameter, which results in a cumbersome $6\times 6$ matrix of secular equation in general case. Note also that the static properties, including the spin-reorientation process as described by Eqs.~(\ref{eqn:angle}),(\ref{eqn:Hsr}), are not affected by the hyperfine coupling.

For the principal in-plane field orientations this problem simplifies and can be solved analytically:

\noindent $\vect{H}||[100]$
\begin{eqnarray}
\omega^2&=&\Delta^2+2\eta m \frac{\omega^2}{\omega^2-\omega_{\rm n}^2}\nonumber\\
\omega^2&=&(\gamma H)^2+\delta^2+2\eta m \frac{\omega^2}{\omega^2-\omega_{\rm n}^2}\label{eqn:hfanis1}
\end{eqnarray}

\noindent $\vect{H}||[110]$, $H>H_{\rm sr}$
\begin{eqnarray}
\omega^2&=&\Delta^2-\frac{\delta^2}{2}+2\eta m \frac{\omega^2}{\omega^2-\omega_{\rm n}^2}\nonumber\\
\omega^2&=&(\gamma H)^2-\delta^2+2\eta m \frac{\omega^2}{\omega^2-\omega_{\rm n}^2} \label{eqn:hfanis2}
\end{eqnarray}

 \noindent $\vect{H}||[110]$, $H<H_{\rm sr}$
\begin{equation}\label{eqn:hfanis3}
    \left|
    \begin{array}{cc}
    2\imath \omega \gamma H{\cal B}&\left[\omega^2-\Delta^2+\delta^2/2-\right.\\
    &\left.-{\cal B}^2 \delta^2- \frac{2 \eta m {\cal C} \omega^2}{\omega^2-\omega_{\rm n}^2}\right]\\
    &\\
    \left[\omega^2-\gamma^2 H^2 \sin 2\varphi-\right.&-2\imath\omega\gamma H{\cal B}\\
    \left.-\delta^2 \cos 4\varphi- \frac{2\eta m \omega^2}{\omega^2-\omega_{\rm n}^2}\right]&
    \end{array}
    \right|=0
\end{equation}

\noindent In the equations above $\Delta$ and $\delta$ are anisotropy-induced gaps introduced in the previous subsection, $m\propto 1/T$ is the equilibrium nuclear magnetization, $\eta=2A\gamma^2/\chi$ is the coupling parameter and $\omega_{\rm n}=\gamma_n H^{\rm (eff)}=(2 A\gamma_n)/\rho$ is the unbiased NMR frequency in the effective field created by electronic spins. In the last equation $\varphi$ is the angle of the order parameter rotation as defined by Eq.~(\ref{eqn:angle}), ${\cal B}=(\cos\varphi-\sin\varphi)/\sqrt{2}$ and ${\cal C}=(\cos\varphi+\sin\varphi)/\sqrt{2}$ for short. The  typical value of unbiased frequency $\omega_{\rm n}/(2\pi)$ in manganese-based antiferromagnets ranges from 400 to 600 MHz \cite{andrienko,turovNMR}. At $\delta=0$ and $\omega\gg\omega_{\rm n}$ the hyperfine gap equals $\sqrt{2\eta m}$, its value in different manganese compounds is of the order of $1...5$~GHz at $T\simeq 1..2$~K \cite{prsos,zaliznyakzorin,andrienkoproz1,andrienkoproz2}.

At high magnetic fields $H>H_{\rm sr}$ and frequencies $\Delta>\omega\gg\omega_{\rm n}$ we obtain for principal in-plane orientations $\vect{H}||[100]$ and $\vect{H}||[110]$

\begin{equation}
\omega^2=(\gamma H)^2+2 \eta m\pm \delta^2.
\end{equation}

 The hyperfine contribution $2\eta m\propto 1/T$ causes the  shift of the low temperature ESR absorption to the left from the paramagnetic one for both field directions solving the contradiction with the experimental data noted in the previous subsection. Our model predicts for fixed frequency resonance fields $H_{\rm res}^{[110]}>H_{\rm res}^{[100]}$ in agreement with the experiment. Now we can quantitatively describe the observed in-plane anisotropy of antiferromagnetic resonance (see Fig.~\ref{fig:angular}). The hyperfine contribution cancels out in the difference of the squares of fixed frequency resonance fields and

 \begin{equation}\label{eqn:Hdiff}
   \left(H_{\rm res}^{[110]}\right)^2-\left(H_{\rm res}^{[100]}\right)^2=2\delta^2/\gamma^2=H_{\rm sr}^2
 \end{equation}

\noindent  this yields another estimate of the spin reorientation field $H_{\rm sr}\approx (0.41\pm0.08)$~kOe. Note, that this estimate tends to underestimate the spin-reorientation field since Eq.~(\ref{eqn:Hdiff}) is derived from $T=0$ model, while at finite temperature deviation of the antiferromagnetic resonance field from the paramagnetic position is smaller.

We can now finalize the determination of hyperfine coupling and in-plane anisotropy parameters from the available data. The spin-reorientation field and the anisotropic contribution to the smaller gap can be estimated from static magnetization \cite{masuda2010,zaliznyak}, ESR line-shape asymmetry (Fig.~\ref{fig:low_freqs}) and ESR resonance field anisotropy (Fig.~\ref{fig:angular}). The hyperfine coupling constant (which is conveniently expressed in terms of unbiased NMR frequency) can be estimated from the temperature dependence of the resonance field (Fig.~\ref{fig:he3(t)} and Eq.~(\ref{eqn:linfit})) and from the values of gaps in ESR $f(H)$ dependencies (Fig.~\ref{fig:f(h)}). The calculation of these parameters relies on the  static susceptibility value $\chi_0$, we used the value for $\vect{H}||[001]$ $\chi_0=0.26$~emu/mole from Ref.~\cite{masuda2010}. The estimation of the unbiased NMR frequency from the gap in ESR $f(H)$ dependence relies on estimation of the anisotropic contribution to the gap $\delta$. The resulting data are summed up in Table  \ref{tab:summary}.

\begin{table}
  \centering
  \caption{Determination of parameters of in-plane anisotropy (at $T=1.8...2$~K) and hyperfine coupling parameters in \bamn{} from different experimental data}\label{tab:summary}
\begin{tabular}{ccc}
&\multicolumn{2}{c}{In-plane anisotropy parameters}\\
&  $H_{\rm sr}$,~kOe&$\delta/(2\pi)$,~GHz\\
\hline
Magnetization \cite{masuda2010}&$1.2\pm0.2$&$3.4\pm 0.6$\\
Magnetization \cite{zaliznyak}&$0.60\pm0.1$&$1.7\pm 0.3$\\
ESR line-shape&&\\
asymmetry&$0.78\pm0.07$&$2.2\pm0.2$\\
Resonance field&&\\
anisotropy at 18~GHz&$0.41\pm0.08$&$1.1\pm0.2$\\
\hline
&\multicolumn{2}{c}{Hyperfine coupling parameters}\\
&  $\omega_{\rm n}/(2\pi)$,~MHz&\\
\hline
Slope of $H^2(1/T)$&$540\pm30$&\\
0.45~K zero-field AFMR  &&\\
gap for $\delta/(2\pi)=3$~GHz& $460\pm50$&\\
($H_{\rm sr}=1.1$~kOe)&&\\
0.45~K zero-field AFMR &&\\
gap for $\delta/(2\pi)=2$~GHz&$530\pm50$&\\
 ($H_{\rm sr}=0.71$~kOe) &&\\
\end{tabular}
\end{table}

All methods yield close estimates of the parameters of anisotropy and hyperfine interaction.  The values found for the unbiased NMR frequency are in the range of the  known values \cite{andrienko,turovNMR}. Slight fine tuning of the parameters can be done in comparison with low-frequency $f(H)$ data (Fig.~\ref{fig:f(h)low}). Softening of the ESR mode at the spin-reorientation transition competes with the hyperfine pulling effect making the position of the resonance absorption at low frequencies very sensitive to the balance between the model parameters. We have found that setting the anisotropic contribution to the smaller gap to $\delta=2$~GHz ($H_{\rm sr}=0.71$~kOe) reproduces better the low frequency $f(H)$ data and also makes different estimates of hyperfine coupling consistent at $\omega_{\rm n}/(2\pi)\approx 530$~MHz. The latter values are considered as the best fit to the full body of our experimental data.

The resulting model $f(H)$ curves calculated for $\vect{H}||[100]$ and $\vect{H}||[110]$  are drawn at Fig.~\ref{fig:f(h)low} in comparison with low-frequency data. The model curves suggest that the microwave  absorption observed at 1.2~GHz and 0.78~GHz turns out to be non-resonant absorption and is most probably due  to the extended wings of the ``true'' resonance curve.

The strength of the hyperfine interaction was also estimated in Ref.~\cite{zaliznyak} from the temperature dependence of the smaller gap in the magnon spectrum as $\widetilde{A}=240$~kOe. This constant is related to the unbiased NMR frequency: $\omega_{\rm n}=\gamma_n \widetilde{A}\langle S \rangle$, which yields $\omega/(2\pi)=590$~MHz.

\subsection{Dipolar and anisotropic contributions to order parameter anisotropy in \bamn{}}

The electron spin resonance experiment confirms easy-plane anisotropy of the N\'{e}el phase of \bamn{}. Within the mean-field approach \cite{goorevich} the  zero-field gap can be expressed via the effective anisotropy field $H_{\rm A}$

\begin{equation}
\Delta=\gamma \sqrt{2 H_{\rm A} H_{\rm E}}=\gamma\sqrt{H_{\rm A} H_\textrm{sat}},
\end{equation}

\noindent
here $H_{\rm E}$ is the exchange field and $H_\textrm{sat}=2 H_{\rm E}$ is the saturation field. This yields an estimate of the anisotropy field $H_{\rm A}\approx 0.85$~kOe.

In the case of uniaxial anisotropy, the anisotropy energy per magnetic ion can be written as  $U_{\rm A} = K \cos^2\Theta$, here $\Theta$ is the angle between the sublattice magnetization and tetragonal axis, and the anisotropy field is $H_{\rm A} = |K|/\mu$, here
$\mu$ is the magnetization of the ion.

The easy-plane anisotropy of the antiferromagnetic order parameter in \bamn{} can originate from two main sources: dipolar interactions and single-ion anisotropy. The dipolar energy of a collinear antiferromagnet  can be calculated straightforwardly in the classical approximation. The dipolar energy for  antiferromagnetic stacked square lattices with inter-layer distance smaller than intra-layer distances favors an easy-plane anisotropy. The dipolar contribution to the anisotropy field (assuming a magnetization of  $4.66\mu_B$  per Mn$^{2+}$ $S=5/2$ ion \cite{zaliznyak}) is 0.34~kOe, which is approximately half of the expected value. Spin-orbital coupling is strongly suppressed for an S-state Mn$^{2+}$ ion. A typical single-ion anisotropy parameter (assuming the notation $D_\textrm{SI} {\widehat{S}_z}^2$) for the  Mn$^{2+}$ ion \cite{altkoz} is $D_\textrm{SI}\simeq 0.6$~GHz which adds approximately 0.5~kOe to the anisotropy field covering the shortage.

There is no in-plane anisotropy of dipolar origin for classical magnetic moments on a square lattice. Lifting of in-plane degeneracy by dipolar interactions via ``order-by-disorder'' mechanism was considered for the antiferromagnet on a cubic lattice \cite{syr2015}. These considerations predict that the cube edges become easy directions. Similar considerations \cite{syr-private} for the tetragonal symmetry of \bamn{} predict that $[110]$-like directions (along the shortest Mn-Mn in-plane bond) should become easy in-plane directions, while neutron scattering \cite{masuda2010} and our experiments indicate that $[100]$-like directions are in-plane easy directions. This means that the choice of the in-plane anisotropy is governed by other anisotropic interactions. Ref.~\cite{zaliznyak} argues that the in-plane anisotropy is influenced by electric polarization of the multiferroic state.

\subsection{ESR linewidth broadening above $T_{\rm N}$: BKT vortices vs critical broadening scenario}
\begin{figure}
\epsfig{clip=, width=\figwidth,file=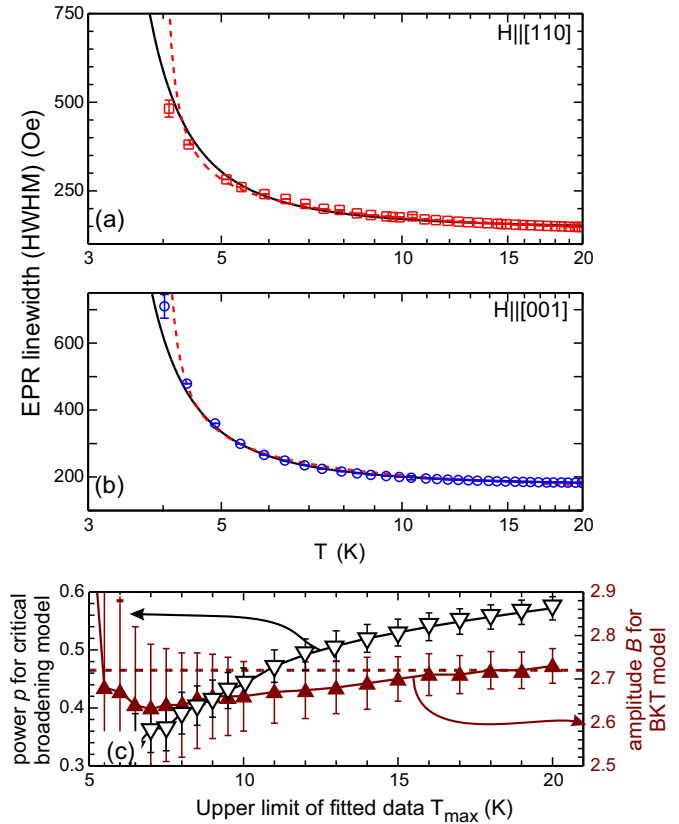}
\caption{(color online) (a) and (b): Temperature dependence of EPR linewidth (symbols) and best fits with critical broadening model (dashed curves) and BKT model (solid curves) for two field directions. Model curves are obtained by fitting the  data in  the range $[4.1; T_\textrm{max}]$ with $T_\textrm{max}=20$~K. (c)  Dependence of model parameters  on temperature interval $[4.1; T_\textrm{max}]$ used for fit. Left y-axis: power $p$ of the critical broadening model (Eq.~(\ref{eqn:crit-theory})) (open triangles). Right y-axis: amplitude $B$ of BKT model with ``ideal'' parameter $b=\pi/2$  (Eq.~(\ref{eqn:BKT-theory})) (filled triangles)}
\label{fig:dH-BKT-vs-crit}
\end{figure}

\begin{table}
\centering
  \caption{Best fit  parameters for critical broadening model and BKT model. Given parameter values are obtained by fitting the data in  the range $[4.1; T_\textrm{max}]$ with $T_\textrm{max}=20$~K.}
  \label{tab:widthfit}
\begin{tabular}{ccc}
model& parameter &value\\
\hline
critical&$H_0^{[001]}$&$(135\pm4)$~Oe\\
broadening&$H_0^{[110]}$&$(115\pm4)$~Oe\\
&$A^{[001]}$&$(76\pm4)$~Oe\\
&$A^{[110]}$&$(89\pm5)$~Oe\\
&$p$&$0.57\pm 0.02$\\
\hline
BKT&$H_0^{[001]}$&$(169\pm2)$~Oe\\
&$H_0^{[110]}$&$(140\pm2)$~Oe\\
&$B$&$(2.72\pm0.04)$~Oe\\
\end{tabular}
\end{table}

The ESR line broadens significantly on cooling below 20~K. Critical broadening of the ESR line close to the transition temperature is quite typical. It can be empirically described as:
\begin{equation}\label{eqn:crit-theory}
\Delta H_\textrm{cr}=\Delta H_0+\frac{A}{(T/T_{\rm N}-1)^p}
\end{equation}

\noindent  Here the offset $\Delta H_0$ models contributions from other relaxation mechanisms, both $\Delta H_0$ and amplitude $A$ are assumed to be anisotropic.

However, since \bamn{} is a quasi-2D system with planar anisotropy it is  subject to the Berezinskii-Kosterlitz-Thouless (BKT) transition model as well. A BKT transition is known to evolve into conventional magnetic ordering once the interactions between the 2D subsystems are present. However, free vortices of a 2D planar magnet contribute to spin dynamics and spin relaxation in the paramagnetic phase. The corresponding contribution to ESR linewidth \cite{kvn} has a characteristic exponential temperature dependence:

\begin{equation}\label{eqn:BKT-theory}
\Delta H_\textrm{BKT}=\Delta H_0+B \exp\left(\frac{3 b}{\sqrt{T/T_\textrm{BKT}-1}}\right)
\end{equation}

\noindent here $b=\pi/2$ for the ideal case (this factor is known to be smaller in some of the real systems), and $T_\textrm{BKT}$ is the BKT transition temperature related to the N\'{e}el temperature as

\begin{equation}\label{eqn:BKT-theory2}
\frac{T_{\rm N}}{T_\textrm{BKT}}-1=\frac{4 b^2}{\left[\ln\left(J/J'\right)\right]^2}.
\end{equation}

\noindent here $J'$ is the inter-planar coupling constant. For the parameters of \bamn{} and $b=\pi/2$ one obtains $T_\textrm{BKT}=2.16$~K.  The anisotropic offset $\Delta H_0$, again, models contributions of other relaxation mechanisms, the amplitude $B$ is isotropic. Such a broadening was observed in quasi-2D magnets on square \cite{jorg}, honeycomb \cite{kvn} or triangular \cite{hemmida1,hemmida2} lattices.

The fit curves using Eqs.~(\ref{eqn:crit-theory}) or (\ref{eqn:BKT-theory}) looks qualitatively similar and differ only quantitatively which complicates unbiased choice between these models. We tried to fit the  X-band (9.40 GHz) data with both models over the temperature range $[4.1~\textrm{K};T_\textrm{max}]$ with $T_\textrm{max}$ varied from 6 to 20~K (see Fig.~\ref{fig:dH-BKT-vs-crit}). The best fit parameters for $T_\textrm{max}=20$~K are shown in Table \ref{tab:widthfit}. Both models follow the experimental data closely (Fig.~\ref{fig:dH-BKT-vs-crit}). However, we have found, that the parameters of the critical broadening model (\ref{eqn:crit-theory}) significantly depend on the choice of the temperature range, while the parameters of the BKT broadening scenario (\ref{eqn:BKT-theory}) are fairly stable (see Fig.~\ref{fig:dH-BKT-vs-crit}). Thus, the description of spin relaxation above the N\'{e}el temperature in terms of the BKT model seems to be more reliable in case of \bamn{}. Note also here that the BKT model fit (\ref{eqn:BKT-theory}) in the case of \bamn{} turns out to be successful assuming the ``ideal case'' value for the parameter $b=\pi/2$ and the BKT temperature $T_\textrm{BKT}$ value determined by the known \cite{masuda2010} exchange coupling constants (\ref{eqn:BKT-theory2}). This reduces number of fitting parameters and further corroborates the validity of the BKT scenario.

\section{Conclusions}
We presented the results of our multi-frequency magnetic resonance study in  the square lattice quasi-2D magnet \bamn{} both above and below the ordering temperature. The antiferromagnetic resonance below the N\'{e}el point is characterized by two zero-field gaps originating from the main easy plane anisotropy  and weaker in-plane anisotropy. The anisotropy effects were found to compete with hyperfine contributions. The analysis of the low-temperature ESR data allowed to determine the hyperfine coupling parameter for this compound. Paramagnetic resonance above the N\'{e}el point demonstrates features related to the vortex dynamics of the planar 2D magnet.

\acknowledgements
The work was supported by the Russian Science Foundation grant 22-12-00259. The authors thank E.Zvereva (Moscow State University) for useful discussion on BKT transition and A.Andrienko (Kurchatov Institute) for discussions on hyperfine effects on ESR, S.Sosin (Kapitza Institute) and A.Smirnov (Kapitza Institute) for numerous helpful discussions.


\begin{thebibliography}{10}

\bibitem{dejong} L. J. de Jongh and A. R. Miedema \emph{``Experiments on simple magnetic model systems''} Advances in Physics, \textbf{23}, 1 (1974) [reprinted as Advances in Physics, \textbf{50}, 947 (2010)].

\bibitem{vasiliev} A. Vasiliev, O. Volkova, E. Zvereva and  M. Markina \emph{``Milestones of low-D quantum magnetism''} npj Quantum Materials \textbf{3}, 18 (2018)

\bibitem{matis} D. C. Mattis \emph{``The Theory of Magnetism Made Simple''} World Scientific Publishing (2006)

\bibitem{bkt} J. M. Kosterlitz and D. J. Thouless \emph{``Ordering, metastability and phase transitions in two-dimensional systems''} Journal of Physics C: Solid State Physics \textbf{6}, 1181 (1973)

\bibitem{QMC-XY-BKT} A. Cuccoli, T. Roscilde, V. Tognetti, R. Vaia, and P. Verrucchi \emph{``Quantum Monte Carlo study of S=1/2 weakly anisotropic antiferromagnets on the square lattice''} Physical Review B \textbf{67}, 104414 (2003)


\bibitem{fortune} N. A. Fortune, S. T. Hannahs, Y. Yoshida, T. E. Sherline, T. Ono, H. Tanaka, and Y. Takano \emph{``Cascade of Magnetic-Field-Induced Quantum Phase Transitions in a Spin-1/2 Triangular-Lattice Antiferromagnet''} Physical Review Letters \textbf{102}, 257201 (2009).

\bibitem{zhitomirskii} M. E. Zhitomirsky and H. Tsunetsugu \emph{``Magnon pairing in quantum spin nematic''}   Europhysics Letters \textbf{92}, 37001 (2010).





\bibitem{mostovoy} S.-W.Cheong and M. Mostovoy \emph{``Multiferroics: a magnetic twist for ferroelectricity''} Nature Materials \textbf{6}, 13 (2007)

\bibitem{ratcliff} W.D. RatcliffII, J.W. Lynn, in \emph{Experimental Methods in the Physical Sciences} \textbf{48} 2015

\bibitem{pyatakov}  A. P. Pyatakov , A. K. Zvezdin \emph{``Magnetoelectric and multiferroic media''} Physics-Uspekhi \textbf{55} 557 (2012)

\bibitem{masuda2010} T. Masuda, S. Kitaoka, S. Takamizawa, N. Metoki, K. Kaneko, K. C. Rule, K. Kiefer, H. Manaka, and H. Nojiri \emph{Instability of magnons in two-dimensional antiferromagnets at high magnetic fields}, Physical Review B \textbf{81}, 100402(R) (2010)

\bibitem{zaliznyak} Shunsuke Hasegawa, Shohei Hayashida, Shinichiro Asai, Masato Matsuura, Zaliznyak Igor, and Takatsugu Masuda \emph{``Nontrivial temperature dependence of magnetic anisotropy in multiferroic Ba$_2$MnGe$_2$O$_7$''} Phys. Rev. Research \textbf{3}, L032023 (2021)

\bibitem{zhitomirskii2} M. E. Zhitomirsky and A. L. Chernyshev \emph{``Instability of Antiferromagnetic Magnons in Strong Fields''} Physical Review Letters \textbf{82}, 4536 (1999)

\bibitem{heeger} A. J. Heeger, A. M. Portis, D. T. Teaney and G. Witt \emph{``Double Resonance and Nuclear Cooling in an Antiferromagnet''} Physical Review Letters \textbf{7}, 307 (1961)

\bibitem{degennes} P. G. de Gennes, P. A. Pincus, F. Hartmann-Boutron, and J. M. Winter \emph{``Nuclear Magnetic Resonance Modes in Magnetic Material. I. Theory''} Physical Review \textbf{129}, 1105 (1963)

\bibitem{witt} G. L. Witt and A. M. Portis \emph{``Nuclear Magnetic Resonance Modes in Magnetic Materials. II. Experiment''} Physical Review \textbf{135}, A1616 (1964)

\bibitem{borovik} A. S. Borovik-Romanov, N. M. Kreines, L. A. Prozorova, \emph{``Antiferromagnetic
resonance in MnCO$_3$``} JETP \textbf{18}, 46 (1964)

\bibitem{andrienko} A.V. Andrienko, V.I. Ozhogin, V.L. Safonov, A.Yu. Yakubovskii \emph{``Nuclear spin wave research''} Sov. Phys. Usp. \textbf{34},  843 (1991)


\bibitem{andrienkoproz1}A.V.Andrienko, L.A.Prozorova, \emph{``Characteristic of the antiferromagnetic resonance spectrum of RbMnCl$_3$''}, Sov.Phys.JETP \textbf{47}, 798 (1978)

\bibitem{andrienkoproz2} A.V.Andrienko, L.A.Prozorova, \emph{``Antiferromagnetic resonance and parametric excitation of spin waves in CsMnCl$_3$''}, , Sov.Phys.JETP \textbf{51}, 1213 (1980)

\bibitem{zaliznyakzorin} I. A. Zaliznyak, N. N. Zorin, and S. V. Petrov \emph{''Investigation of a gap in the AFMR spectrum in the
quasi-one-dimensional hexagonal antiferromagnet CsMnBr$_3$''} JETP Letters  \textbf{64}, 473 (1996) [Pis'ma Zh. Eksp. Teor. Fiz. \textbf{64},  433 (1996)]

\bibitem{prsos} L. A. Prozorova, S. S. Sosin, D. V. Efremov and  S. V. Petrov \emph{``Investigation of the hyperfine interaction in the antiferromagnetic CsMnI$_3$''} Journal of Experimental and Theoretical Physics \textbf{85}, 1035 (1997) [Zh.Exp.Teor.Fiz. \textbf{112}, 1893 (1997)]

\bibitem{murakawa} H. Murakawa, Y.Onose, S.Miyahara, N.Furukawa, Y.Tokura \emph{``Compehensive study of the ferroelectricity induced by the spin-dependent d-p hybridization mechanism in Ba$_2$XGe$_2$O$_7$ (X=Mn, Co, Cu)''} Physical Review B \textbf{85}, 174106 (2012)

\bibitem{sazonov} A. Sazonov , V. Hutanu, M. Meven, G. Roth, R.
Georgii, T. Masuda and B\'{a}lint N\'{a}fr\'{a}di \emph{``Crystal Structure of Magnetoelectric Ba$_2$MnGe$_2$O$_7$ at Room and Low Temperatures by Neutron Diffraction''} Inorg. Chem. \textbf{57}, 5089 (2018)

\bibitem{iguchi} Y. Iguchi, Y. Nii, M. Kawano, H. Murakawa, N. Hanasaki and Y. Onose \emph{``Microwave nonreciprocity of magnon excitations in the noncentrosymmetric antiferromagnet Ba$_2$MnGe$_2$O$_7$''}, Physical Review B \textbf{98}, 064416 (2018)

\bibitem{goorevich} A. G. Gurevich and G. A. Melkov, \emph{``Magnetic oscillations and waves''}, 464 pp., CRC Press, London (1996)

\bibitem{kubo} T. Nagamiya, K. Yosida and R. Kubo \emph{``Antiferromagnetism''} Advances in
Physics, \textbf{4}, 1 (1955)

\bibitem{kolesik} M. Kolesik, M. Suzuki \emph{Accurate estimates of 3D Ising critical exponents using the coherent-anomaly method} Physica A: Statistical Mechanics and its Applications, \textbf{215}, 138 (1995)

\bibitem{campostrini} M. Campostrini, M. Hasenbusch, A. Pelissetto, P. Rossi, E. Vicari \emph{Critical behavior of the three-dimensional XY universality class}, Physical Review B \textbf{63}, 214503 (2001)

\bibitem{andrmar} A. F. Andreev and V. I. Marchenko \emph{``Symmetry and the macroscopic dynamics of magnetic materials''}  Usp. Fiz. Nauk \textbf{130}, 39 (1980)  [Sov. Phys. Usp. \textbf{23} 21 (1980)].
\bibitem{udalov} O.G. Udalov \emph{``NMR spectrum in a Mn$_3$Al$_2$Ge$_3$O$_12$ noncollinear antiferromagnet''} Journal of Experimental and Theoretical Physics \textbf{113}, 490 (2011) [Zh.Exp.Teor.Fiz. \textbf{140}, 561 (2011)]

\bibitem{turovNMR} E.A. Turov, M.P. Petrov \emph{``Nuclear magnetic resonance in ferro- and antiferromagnets''},  Israel Program for Scientific Translations; First Edition (1972) [Russian version: Nauka (1969)]





\bibitem{altkoz} S. A. Altshuler, B. M. Kozyrev \emph{``Electron Paramagnetic Resonance in Compounds of Transition Elements''}, Nauka, Moscow (1972) [English version:  S. A. Al'tshuler,  B. M. Kozyrev, John Wiley and Sons; 2nd edition (1974)]



\bibitem{syr2015} L. A. Batalov and A. V. Syromyatnikov \emph{``Breakdown of long-wavelength magnons in cubic antiferromagnets with dipolar forces at small temperature''} Phys. Rev. B 91, 224432 (2015)

\bibitem{syr-private} A. V. Syromyatnikov, private communications

\bibitem{jorg} T. F\"{o}rster, F. A. Garcia, T. Gruner, E. E. Kaul, B. Schmidt, C. Geibel, and J. Sichelschmidt \emph{``Spin fluctuations with two-dimensional XY behavior in a frustrated $S = 1/2$ square-lattice ferromagnet''} Phys.Rev. B \textbf{87}, 180401(R) (2013)

\bibitem{kvn} M. Heinrich, H.-A. Krug von Nidda, A. Loidl, N. Rogado, and R. J. Cava \emph{\emph{``Potential Signature of a Kosterlitz-Thouless Transition in BaNi$_2$V$_2$O$_8$''}}  Phys. Rev. Lett. \textbf{91}, 137601 (2003)

\bibitem{hemmida1} Mamoun Hemmida, Hans-Albrecht Krug von Nidda, and Alois Loidl \emph{``Traces of Z$_2$-Vortices in CuCrO$_2$, AgCrO$_2$, and PdCrO$_2$''} Journal of the Physical Society of Japan \textbf{80}, 053707  (2011)

\bibitem{hemmida2} M. Hemmida, H.-A. Krug von Nidda, N. B\"{u}ttgen, A. Loidl, L. K. Alexander, R. Nath, A. V. Mahajan,
R. F. Berger, R. J. Cava, Yogesh Singh, and D. C. Johnston \emph{``Vortex dynamics and frustration in two-dimensional triangular chromium lattices''} Phys. Rev. B \textbf{80}, 054406 (2009)

\end{thebibliography}
\end{document}